\documentclass[aps,prl,twocolumn,groupedaddress]{revtex4-1}
\usepackage{amsmath}
\usepackage{graphicx,color}% Include figure files
\usepackage{dcolumn}% Align table columns on decimal point
\usepackage{bm}% bold math
\usepackage{CJK}
\usepackage{mathrsfs}

\newcommand{\nc}{\newcommand}       % new command
\nc{\vc}[1] {\mbox{\boldmath $#1$}} % boldmath(vector)
\nc{\del}       {\partial}              % bra state
\nc{\bra}       {\langle}               % bra state
\nc{\ket}       {\rangle}               % ket state
\nc{\bras}[1]   {\langle #1|}           % bra state
\nc{\kets}[1]   {|#1\rangle}            % ket state
\nc{\mapleft}[1]{           % something under arrow
 \smash{\mathop{\,          %
  \hbox to 1.5cm{\rightarrowfill}\, }\limits_{#1}}}
\nc{\beq}     {\begin{eqnarray}} \nc{\eeq}    {\end{eqnarray}}
\nc{\nn}      {\\\nonumber} \nc{\vs}      {\vspace{-0.275cm}}
\nc{\fra}    {\frac{1}{2}}

\begin{document}
%\begin{CJK}{UTF8}{song}

%Title of paper
\title{Relativistic Brueckner-Hartree-Fock theory for finite nuclei}

\author{Shihang Shen}% (ÉêʱÐÐ)}
 \affiliation{State Key Laboratory of Nuclear Physics and Technology, School of Physics, Peking University, Beijing 100871, China}
 \affiliation{RIKEN Nishina Center, Wako 351-0198, Japan}

\author{Jinniu Hu}% (ºú½ðÅ£)}
 \affiliation{Department of Physics, Nankai University, Tianjin 300071, China}

\author{Haozhao Liang}% (ÁººÀÕ×)\footnote{Email: haozhao.liang@riken.jp}}
 \affiliation{RIKEN Nishina Center, Wako 351-0198, Japan}
 \affiliation{Department of Physics, Graduate School of Science, The University of Tokyo, Tokyo 113-0033, Japan}

\author{Jie Meng}% (ÃϽÜ)\footnote{Email: mengj@pku.edu.cn}}
 \affiliation{State Key Laboratory of Nuclear Physics and Technology, School of Physics,
              Peking University, Beijing 100871, China}
 \affiliation{School of Physics and Nuclear Energy Engineering, Beihang University,
              Beijing 100191, China}
 \affiliation{Department of Physics, University of Stellenbosch, Stellenbosch, South Africa}

\author{Peter Ring}
 \affiliation{State Key Laboratory of Nuclear Physics and Technology, School of Physics,
              Peking University, Beijing 100871, China}
 \affiliation{Physik-Department der Technischen Universit\"at M\"unchen, D-85748 Garching, Germany}

\author{Shuangquan Zhang}% (ÕÅË«È«}
 \affiliation{State Key Laboratory of Nuclear Physics and Technology, School of Physics,
              Peking University, Beijing 100871, China}

\date{\today}

\begin{abstract}
Starting with a bare nucleon-nucleon interaction, for the first time the full relativistic
Brueckner-Hartree-Fock equations are solved for finite nuclei in a Dirac-Woods-Saxon basis.
No free parameters are introduced to calculate the ground-state properties of finite nuclei. The nucleus $^{16}$O is investigated as an example. The resulting ground-state properties, such as binding energy and charge radius, are considerably improved as compared with the non-relativistic Brueckner-Hartree-Fock results and much closer to the experimental data. This opens the door for \emph{ab initio} covariant investigations of heavy nuclei.
\end{abstract}

% PACS, the Physics and Astronomy
                             % Classification Scheme.
%\keywords{Suggested keywords}%Use showkeys class option if keyword
                              %display desired

%\maketitle must follow title, authors, abstract, \pacs, and \keywords
\maketitle

% body of paper here - Use proper section commands
% References should be done using the \cite, \ref, and \label commands
%\section{Introduction}
% Put \label in argument of \section for cross-referencing
%\section{\label{}}

\textit{Ab initio} calculations, i.e., the proper description of finite nuclei with a bare nucleon-nucleon ($NN$) interaction adjusted only to the scattering data of free nucleons, form a central problem of theoretical nuclear physics since the middle of last century. These realistic $NN$ interactions are characterized by a repulsive core at short distance~\cite{Jastrow1951_PR81-165}, a strong attraction at intermediate range, and are dominated by one-pion exchange at large distance~\cite{Machleidt1989_ANP19-189}. Many methods have been proposed in the past to treat their singular behavior, such as Brueckner theory~\cite{Brueckner1954_PR95-217,Day1967_RMP39-719} and variational methods~\cite{Jastrow1955_PR98-1479,Day1978_RMP50-495}. Recently with the great progress of the high-precision $NN$ interactions, such as Reid93~\cite{Stoks1994_PRC49-2950}, AV18~\cite{Wiringa1995_PRC51-38}, CD Bonn~\cite{Machleidt2001_PRC63-24001}, and chiral potentials~\cite{Epelbaum2009_RMP81-1773,Machleidt2011_PR503-1},  and with increasing computer technology, more and more \textit{ab initio} methods have been developed to study the nuclear many-body system, e.g., the Green's function Monte Carlo method~\cite{Carlson2015_RMP87-1067}, the self-consistent Green's function method~\cite{Dickhoff2004_PPNP52-377}, the coupled-cluster method~\cite{Hagen2014_RPP77-096302}, the lattice chiral effective field theory~\cite{LEE-D2009_PPNP63-117}, and the no-core shell model~\cite{Barrett2013_PPNP69-131}.

It has been found rather early that all the non-relativistic potentials systematically failed to reproduce the saturation properties of infinite nuclear matter in \textit{ab initio} calculations. The saturation binding energies and corresponding densities obtained with various forces are distributed on the so-called Coester line~\cite{Coester1970_PRC1-769}, which bypasses the experimental area considerably. Therefore, it has been concluded that the bare three-body force plays an essential role in the nuclear many-body problem~\cite{Fujita1957_PTP17-360,Brown1969_NPA137-1}, and all the modern non-relativistic investigations use such phenomenological terms. In this way, one was able to reproduce the saturation properties of nuclear matter~\cite{SONG-HQ1998_PRL81-1584} and the ground-states and a few excited states of light nuclei~\cite{Pieper2001_ARNPS51-53}. However, these calculations require extreme computational efforts and are very difficult to extend for heavy nuclei.

On the other side it is known since more than 30 years that relativistic Brueckner-Hartree-Fock (RBHF)~\cite{Anastasio1983_PR100-327,Brockmann1984_PLB149-283,terHaar1987_PR149-207} theory has no problems to reproduce empirical saturation data, even without three-body forces. This method is based on relativistic Hartree-Fock calculations with an effective interaction, the $G$-matrix~\cite{Brueckner1954_PR95-217}. It describes the scattering of two particles in the nuclear medium and is derived from the solution of the Bethe-Goldstone equation~\cite{Bethe1957_PRSA238-551}. It depends, in a self-consistent way, on the density and therefore, in finite nuclei, on the position of the interacting particles.
Non-relativistic BHF theory has been applied to study finite medium mass nuclei in the 1970's~\cite{Becker1974_PRC9-1221} with different $NN$ interactions. Because of the missing three-body interaction, however, these applications have not been very successful.

Inspired by the success of relativistic Brueckner theory in nuclear matter~\cite{Anastasio1983_PR100-327,Brockmann1984_PLB149-283,terHaar1987_PR149-207}, it is a natural extension to study finite nuclei in the same framework. Nevertheless, in the eighties, this was a formidable task and different approximations have been adopted. M\"uther, Machleidt, and Brockmann applied the effective density approximation (EDA)~\cite{Muether1988_PLB202-483,Muether1990_PRC42-1981}, where non-relativistic BHF equations were solved and relativistic effects were taken into account using the effective mass derived from from RBHF calculation in nuclear matter. Later on, the local density approximation (LDA) was introduced by mapping the density-dependence of the $G$-matrix in nuclear matter to effective interactions used in density-dependent relativistic Hartree and Hartree-Fock (DDRH/DDRHF) calculations (see Refs.~\cite{Brockmann1992_PRL68-3408,Fritz1993_PRL71-46,VanDalen2010_IJMPE19-2077} and references therein). However, as discussed in detail in Ref.~\cite{Giai2010_JPG37-064043} this mapping is not unique and therefore different approximations lead to rather different results. A self-consistent study for finite nuclei is highly demanded. See the Supplemental Material for the different results obtained within these frameworks.

In this Letter, we report the first self-consistent RBHF study for finite nuclei with a realistic $NN$ interaction, avoiding the EDA and LDA approximations. The Bethe-Goldstone (BG) and relativistic Hartree-Fock (RHF) equations are solved simultaneously within a Dirac-Woods-Saxon (DWS) basis~\cite{ZHOU-SG2003_PRC68-034323}, a basis of the eigenfunctions of a Dirac equation with scalar and vector potentials of Woods-Saxon type in a spherical box of finite size.
%\textcolor{blue}{The ground-state properties of $^{16}$O are studied in detail using the realistic interaction %Bonn A~\cite{Machleidt1989_ANP19-189}.}

%\section{Theoretical framework}
Hartree-Fock theory in finite nuclei~\cite{Ring1980} can be solved in $r$-space or in a discrete basis space characterized by the quantum numbers $i,j$:
\begin{equation}
\label{eq:HF}
\sum_{i'}\left(t_{ii'}+\sum_{jj'}\bar{V}_{iji'j'}\rho_{j'j}\right)D_{i'a}=\varepsilon_{a}D_{ia}.
\end{equation}
Here $D_{ia}$ are the expansion coefficients of the eigenstate $|a\rangle=\sum_i D_{ia}|i\rangle$ in terms of the DWS basis states $|i\rangle$. $t=\bm{\alpha}\cdot\mathbf{p}+\beta M$ is the kinetic energy in the relativistic case,
$\bar{V}_{iji'j'}=\langle ij|V|i'j'-j'i'\rangle$ is the anti-symmetrized matrix element of the interaction $V$, and $\rho_{j'j}$ is the density matrix element related to the expansion coefficients:
\begin{equation}
\label{density}
\rho_{j'j}=\langle a^\dag_j a^{}_{j'}\rangle = \sum^A_{c=1}D^{}_{j'c}D^*_{jc}.
\end{equation}
where the index $c$ runs over all the states in the Fermi sea ($\varepsilon_c\leq\varepsilon_F$). In the relativistic case the vacuum polarization, i.e. the sum over the states in the Dirac sea, is usually neglected in the framework of the {\it no-sea} approximation.

In Brueckner-Hartree-Fock theory, the interaction $V$ in Eq.~(\ref{eq:HF}) is replaced by the $G$-matrix. It describes the scattering of two nucleons in the nuclear medium and depends on the starting energy $W$, i.e. on the energy of the two incoming particles. The $G$-matrix sums up all the ladder diagrams with two particles in the intermediate states and is deduced from the Bethe-Goldstone equation~\cite{Brueckner1954_PR95-217,Bethe1971_ARNS21-93,Krenciglowa1976_ANNY101-154}:
\begin{equation}\label{eq:BG}
\bar{G}_{aba'b'}(W)=\bar{V}_{aba'b'}+\frac{1}{2}\sum_{mn}
\frac{\bar{V}_{abmn}\,\bar{G}_{mna'b'}(W)}
{W-\varepsilon_m-\varepsilon_n}.
\end{equation}
Here $V$ is the bare nucleon-nucleon interaction and the intermediate states $m,n$ run over all states above the Fermi surface with ($\varepsilon_m,\varepsilon_n>\varepsilon_F$). This involves the calculation of the Pauli operator
\begin{equation}
Q_{F}=\frac{1}{2}\sum_{m,n>F}
|mn\rangle\langle{mn}|.
\label{Pauli}
\end{equation}
The single-particle energies $\varepsilon_a$ in the denominator of the BG-equation (\ref{eq:BG}) and the
single-particle wave functions (expressed by the coefficients $D_{ia}$) in the Pauli operator are determined by
the equation of motion for the nucleons, i.e. by the solution of the RHF-equation
\begin{equation}\label{eq:RHF}
\sum_{a'}\left(t_{aa'}+\sum_{bb'}\bar{G}_{aba'b'}\rho_{b'b}\right)D_{a'k}=\varepsilon_{k}D_{ak}.
\end{equation}
It depends on the $G$-matrix and therefore on the solution of the BG-equation (\ref{eq:BG}). Therefore the BG-equation (\ref{eq:BG}) and the RHF-equations (\ref{eq:RHF}) form a complicated system of coupled equations which has to be solved by iteration. After convergence we have a self-consistent solution with single-particle energies and wave functions. We are thus able to evaluate the ground-state energy of the finite nucleus
\begin{equation}
E_{\text{RBHF}}=\sum^A_{c=1}\varepsilon_c-\frac{1}{2}\sum^A_{c,c'=1}\bar{G}_{cc'cc'},
\end{equation}
where the indices $c$ and $c'$ run over all the occupied states in the Fermi sea of the self-consistent potential. We also can calculate the expectation values of single-particle operators such as nuclear rms radii $\langle r^2\rangle^{1/2}$ for protons and neutrons.

We solve the RHF equations (\ref{eq:RHF}) in a Dirac-Woods-Saxon basis~\cite{ZHOU-SG2003_PRC68-034323}. We confirmed the validity of this RHF code in the DWS basis by comparing its results for the ground-state properties of $^{16}$O with those of other RHF-codes in the oscillator basis~\cite{Lalazissis2009_PRC80-041301} and in coordinate space~\cite{Long2006_PLB640-150}.

The BG equation~(\ref{eq:BG}) is solved by matrix inversion~\cite{Haftel1970_NPA158-1} in the space of pair states $|ab\rangle$. These are pairs of Dirac-spinors with the full relativistic structure coupled
to good angular momentum $J$ (particle-particle ($pp$) coupling). The indices $a$ and $b$ run over
all solutions of the Dirac equation (with positive and negative energies). The
BG-equations are solved for each value of $J$. This leads, for the various
$J$-values, to a set of $pp$-coupled matrix elements of the $G$-matrix. The
particle-hole ($ph$) coupled matrix elements of the $G$-matrix with $I=0$ needed for the
solution of the RHF equation (\ref{eq:RHF}) in the next step of the iteration
are obtained by recoupling. The Pauli operator (\ref{Pauli}) and all its
relativistic structure is here fully taken into account. In particular there
is no angle averaging \cite{Schiller1999_PRC59-2934,Suzuki2000_NPA665-92}
involved as it is the case in most Brueckner calculations in nuclear matter.
Of course, in agreement with the \textit{no-sea approximation}, the sum of the
intermediate states $|mn\rangle$ in Eq. (\ref{eq:BG}) does not include
scattering to states in the Dirac sea.

In practice there are several problems. First, the $G$-matrix used in the
RHF-equation (\ref{eq:RHF}) depends on the energy. As it turns out this energy
dependence is not uniquely defined. Several choices have been used in the
literature ~\cite{Rajaraman1967_RMP39-745,Bethe1971_ARNS21-93}. We use here
the so-called \textit{continuous choice} for the evaluation of the mean
potential as described in Eq. (1) of Ref.~\cite{Davies1969_PRC177-1519}. The
BG equation~(\ref{eq:BG}) is solved for four different starting energies, and
the $G$-matrix elements for the various starting energies required in this
choice are obtained by interpolation~\cite{Davies1969_PRC177-1519}.

Second, the Pauli operator in Eq. (\ref{eq:BG}) is obviously defined in the
basis, in which the Dirac field of RHF equation (\ref{eq:RHF}) is diagonal.
This requires, in each step of the iteration, a transformation of the basis
and is connected with a considerable numerical effort. Therefore we adopt in
this investigation a further approximation: assuming that the wave functions
in the Pauli operator (\ref{Pauli}) are close to those of the original DWS
basis, we evaluate the BG equation in this fixed basis. In the denominator of
the BG equation~(\ref{eq:BG}), however, we have the starting
energy $W$, i.e. the energy of the two incoming particles before the
scattering process and the energy $\varepsilon_{m}+\varepsilon_{m}$ of the
particles afterwards. We use in all these cases the self-consistent single-particle
energies of the corresponding RHF-potential with the same quantum
numbers ($\ell j$) and with the same number of radial nodes.

As an application we consider the nucleus $^{16}$O. We use the realistic
$NN$ interaction Bonn A which has been adjusted to the $NN$
scattering data in Ref.~\cite{Machleidt1989_ANP19-189}. This is a relativistic
two-body potential based on the exchange of the six mesons $\sigma$, $\omega$,
$\rho$, $\pi$, $\eta$, and $\delta$, whose masses are less than $1$ GeV with
monopole form factors. The potential for the DWS basis is taken from
Ref.~\cite{Koepf1991_NPA533-95}. In addition, the Coulomb exchange term is taken into account through the
relativistic local density approximation~of Ref. \cite{GU-HQ2013_PRC87-041301}.
The microscopic center-of-mass correction $-{\bm P}^{2}/2AM$ is
included in the total energy and, as in the density functional SLy6 of Ref.
\cite{Chabanat1998_NPA635-231}, it has been taken into account in the iterative solution
of the RHF equation.

\begin{figure}[!htb]
\includegraphics[width=8cm]{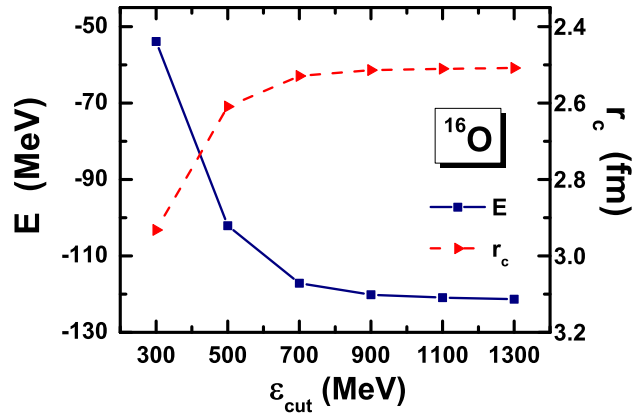}
\caption{(Color online) Total energy $E$ and charge radius $r_c$ of $^{16}$O as a function of energy cut-off $\varepsilon_{\rm cut}$ calculated within RBHF theory with the realistic $NN$ interaction Bonn A~\cite{Machleidt1989_ANP19-189}.
\label{fig1}}
\end{figure}

It is well known that the bare $NN$ force contains a strong tensor part
connecting the nucleons in the Fermi sea to states with high momenta in the
continuum. In order to take this coupling fully into account one needs a
relatively large basis and it is crucial to investigate the convergence of the
RBHF calculation with the size of this basis. Therefore we show in
Fig.~\ref{fig1} the resulting total energy and charge radius of $^{16}$O as
functions of the cut-off energy $\varepsilon_{cut}$. The cut-off energy means
that for each ($\ell j)$ value all the single-particle states with
$\varepsilon_{i}\leq\varepsilon_{cut}$ are taken into account in the DWS basis.
With increasing $\varepsilon_{cut}$ the energy and the charge radius of
$^{16}$O decrease. We find satisfying convergence for $\varepsilon_{cut}=1.1$
GeV with $E=-120.7$ MeV and $r_{c}=2.52$ fm. Therefore, in the following, this
value will be used.

The ground-state properties of $^{16}$O are listed in Table \ref{tab1}: the
total energy $E$, the charge radius $r_c$, the point matter radius $r_m$,
and the proton spin-orbit splitting for the $1p$ shell.
The results of our full RBHF calculation are compared with the corresponding experimental
data~\cite{WANG-M2012_CPC36-1603,Angeli2013_ADNDT99-69,Coraggio2003_PRC68-034320}
and with several other calculations: BHF is a non-relativistic Brueckner
calculation~\cite{Muether1990_PRC42-1981} based on the interaction Bonn A. We
also show results obtained in RHF-calculations with the phenomenological
effective interaction PKO1~\cite{Long2006_PLB640-150}, which has been fitted to
binding energies and charge radii of a set of spherical nuclei. It is seen that the
ground-state properties in RBHF theory are improved considerably as compared
with the non-relativistic results. The deviation from the experimental values
have been decreased from 18\% to 6\% in the case of the energy and from 16 \%
to 7 \% in the case of the charge radius, which is consistent with the
conclusions in the infinite nuclear matter~\cite{Brockmann1990_PRC42-1965}.
This energy of $^{16}$O is also very close to the value of $E=-119.7$
MeV obtained within the No Core Shell Model (NCSM) using the chiral $NN$
interaction N$^{3}$LO~\cite{Roth2011_PRL107-072501} and to the value of $E=-121.0$
MeV obtained within the Coupled Cluster (CC) method~\cite{Hagen2009_PRC80-021306}. The spin-orbit
splittings in RBHF theory is only slightly smaller than the experimental data.
For the $1p$ proton shell we have a deviation of about $5\%$. Of course the
results of the calculations with PKO1 which has been fitted to these data
shows only a very small deviation of 0.5 \% for the energy, of 1.6 \% for the
radius and the spin-orbit splitting.

%%%%%%%%%%%%%%%%%%%%%%%%%%%%%%%%%%%%%%%%%%%%%%%%%%%%%%%%%%%%%%%%%%
%%%%%%%%%%%%%%%%%%%%%%  Table I %%%%%%%%%%%%%%%%%%%%%%%%%%%%%%%%%%
%%%%%%%%%%%%%%%%%%%%%%%%%%%%%%%%%%%%%%%%%%%%%%%%%%%%%%%%%%%%%%%%%%
\begin{table}%[t]
\centering
\begin{ruledtabular}
\begin{tabular}{lcccc} % cD{.}{.}{4} cD{.}{.}{3} cD{.}{.}{2} cD{.}{.}{1} r@{.}l r@{.}l}
&\multicolumn{1}{c}{$E$ (MeV)} & \multicolumn{1}{c}{$r_c$ (fm)} & \multicolumn{1}{c}{$r_m$ (fm)} &
\multicolumn{1}{c}{$\Delta E_{\pi1p}^{ls}$ (MeV)} \\
\hline
Exp. \cite{WANG-M2012_CPC36-1603,Angeli2013_ADNDT99-69,Ozawa2001_NPA691-599,Coraggio2003_PRC68-034320} &
$-127.6$ & $2.70$ & $2.54$ & $6.3$ \\
RBHF & $-120$.$7$ & $2.52$ & $2.38$ & $6$.$0$ \\
BHF \cite{Muether1990_PRC42-1981} & $-105.0$ & $2.29$ & $-$ & $7.5$ \\
DDRH \cite{VanDalen2011_PRC84-024320}& $-106.4$ & $2.72$ & $-$ &  $-$      \\
DDRHF \cite{VanDalen2011_PRC84-024320} & $-142.6$ & $2.62$ & $-$ & $4.5$ \\
NCSM \cite{Roth2011_PRL107-072501}  & $-119.7$ &  $-$ & $-$ & $-$ \\
CC \cite{Hagen2009_PRC80-021306}  & $-121.0$ & $-$ & 2.30 & $-$ \\
PKO1 \cite{Long2006_PLB640-150}  & $-128.3$ & $2.68$ & 2.54 & $6.4$ \\
\end{tabular}
\end{ruledtabular}
\caption{Energy, charge radius, point matter radius, and $\pi1p$ spin-orbit splitting in $^{16}$O with the interaction Bonn A in RBHF, BHF \cite{Muether1990_PRC42-1981}, in RBHF with LDA (DDRH and DDRHF) \cite{VanDalen2011_PRC84-024320}; with N$^3$LO in NCSM \cite{Roth2011_PRL107-072501} and the Couple-Cluster method (CC) \cite{Hagen2009_PRC80-021306}; with PKO1 in CDFT \cite{Long2006_PLB640-150}.}\label{tab1}
\end{table}

\begin{figure}
\includegraphics[width=8cm]{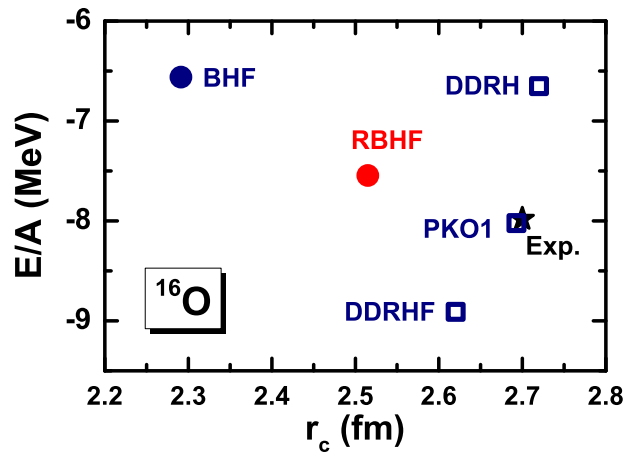}
\caption{(Color online) Energy per particle and charge radius of $^{16}$O by (relativistic) BHF theories compared with experimental data and other calculations. See text for details.
\label{fig2}}
\end{figure}

Next we compare in Fig.~\ref{fig2} our self-consistent results in finite nuclei with
those obtained in Ref.~\cite{VanDalen2011_PRC84-024320} by two
"\textit{ab initio}" calculations  based on the LDA. There the full
RBHF equations are solved for nuclear matter at various densities and
the corresponding scalar and vector self-energies are derived.
Then density-dependent coupling strengths for the
exchange of various mesons in a relativistic Hartree (DDRH) or a Hartree-Fock (DDRHF) model
have been adjusted to these results. In this case it is possible to investigate
finite nuclei in an \textit{ab inito} approach without any
phenomenological parameters.

The results of these calculations based on the LDA also listed in the 4$^{th}$ and
the 5$^{th}$ rows of Table \ref{tab1}. The charge radii are rather well
reproduced in these local density approaches. These values are considerably
closer to the data than our calculations. There is, however, a
considerable difference in the total energy between Hartree and
Hartree-Fock results. DDRH shows 17 \% underbinding as compared with the
experimental value. The Fock-term overshoots and shows a 12 \% overbinding. We
have to keep in mind, however, that the methods connected with the LDA are
always based on a mapping and there exist important uncertainties in this
mapping, which deserve further investigations. The full solution of the RBHF
equations in finite nuclei presented here is therefore very essential for
comparisons in the future.

\begin{figure}
\includegraphics[width=7.5cm]{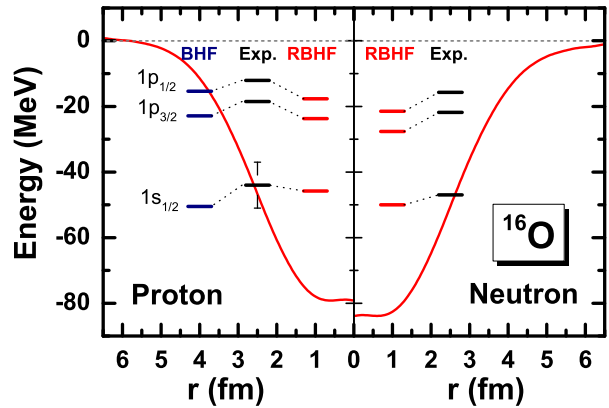}
\caption{(Color online) Single-particle spectra for protons and neutrons obtained from the solution of the RHF-equation~(\ref{eq:RHF}) are compared with experimental values~\cite{Coraggio2003_PRC68-034320}. The red line corresponds to a local potential $V(r)+S(r)$ in a Dirac equation producing the same wave function and the same eigenvalue for the lowest $1s_{1/2}$ state as the full RBHF-equation.
    \label{fig3}}
\end{figure}

The single-particle energy levels of $^{16}$O for protons and neutrons obtained
from the full RBHF calculation are plotted in Fig.~\ref{fig3}. They are
compared with experimental data and (for the protons) with the results of
non-relativistic BHF calculations. It is clearly seen, that for protons and
neutrons the $p$-levels are slightly too low as compared with the data.
This can possibly be understood by the fact, that in all density-dependent
mean-field calculations one finds a too large gap in the single-particle
spectrum at the Fermi surface. This seems to apply also for our results. In
the present method we sum up only ladder diagrams and neglect more complicated
configurations in the intermediate states, which lead to energy dependent
self-energies and shifts of the single-particle spectrum in the neighborhood
of the Fermi surface~\cite{Litvinova2006_PRC73-044328}.

In summary, the full relativistic Brueckner-Hartree-Fock (RBHF) equations have
been solved for the first time for finite nuclei in a Dirac-Woods-Saxon (DWS) basis.
In each step of the iteration the matrix elements of the $G$-matrix describing
the scattering in the finite nuclear medium are determined by the solution of
the Bethe-Goldstone equations in the DWS basis of sufficient size with the
single-particle energies of the corresponding self-consistent relativistic
potentials. The relativistic structure of the two-body matrix elements as well
as of the Pauli operator is fully taken into account. The only input is the
bare $NN$-interaction Bonn A adjusted to the scattering phase shifts in
Ref.~\cite{Machleidt1989_ANP19-189}. No other parameter is used. Since
nuclear matter calculations within the same framework produce results far away from
the Coester line and close to the experimental values of saturation, we
neglect at this stage three-body forces.

Modern non-relativistic BHF-calculations in nuclear matter include three-body forces and
are therefore able to reproduce the right saturation properties~\cite{Baldo1997}.
An investigation of the microscopic origin of the various three-body terms~\cite{Zuo2002_NPA706-418} showed that below and around nuclear saturation density the three-body diagram with a $N\bar{N}$-pair in the intermediate state that is coupled by $\sigma$-mesons to the other two nucleons (the so-called Z-diagram), is able to reproduce to a large extent the total contribution of three-body forces. The other three-body diagrams cancel more or less in this density region. Therefore the successful description of nuclear saturation in relativistic Brueckner Theory is understandable, because it includes automatically the Z-diagram~\cite{Brown1987_CommNPP17-39,Zuo2002_NPA706-418,Giai2010_JPG37-064043}. At the moment it is an open question, to what extend one needs in a relativistic description additional bare three-body forces.

As an example we consider the nucleus $^{16}$O. We find convergence for a
cut-off in the single-particle energy at $\varepsilon_{\text{cut}}\approx1.1$
GeV.  The resulting binding energy is in good agreement with state-of-the-art
parameter-free \textit{ab initio} calculations, but it still deviates by 6 \%
from the experimental value and the charge radius agrees with the experimental
value up to 7 \%. Also the spin-orbit splitting is well reproduced.

Despite the good agreement of these results, there is room for improvements.
The RBHF-theory presented here is no exact solution of the nuclear many-body problem.
So far, rearrangement terms are not taken into account and higher order
diagrams in the hole-line expansion are not included. Those effects have been
taken into account in some approximation in non-relativistic
calculations~\cite{Negele1982_RMP54-913}, but for relativistic theories
they are left for future investigations.

On the other side, our method has the potential to investigate
heavier nuclei, where exact solutions are impossible, in particular
systems without spin saturation and with large neutron excess. In this
case we hope to be able to gain a parameter-free, microscopic understanding
of open questions in modern phenomenological density functional theories,
such as their isospin dependence or the importance of the tensor
terms~\cite{Lalazissis2009_PRC80-041301}.

\begin{acknowledgments}
We thank Wenhui Long for the discussions and for providing the RHF code.
This work was partly supported by the Major State 973 Program of China No.~2013CB834400, Natural Science Foundation of China under Grants No.~11175002, No.~11335002, No.~11405090, No.~11375015, and No.~11621131001, the Research Fund for the Doctoral Program of Higher Education under Grant No.~20110001110087, the DFG cluster of excellence \textquotedblleft Origin and Structure of the Universe\textquotedblright\ (www.universe-cluster.de), the CPSC Grant No. 2012M520100, and the RIKEN IPA and iTHES projects.
\end{acknowledgments}

\bibliographystyle{C:/Users/Peter/Dropbox/Programme/Jabref/prsty}
%\bibliography{RBHF}
%\bibliography{C:/Users/Peter/Dropbox/Programme/Jabref/Ring}

%\end{CJK}
\end{document}